\documentclass[aps,amssymb,amsmathprx,reprint,preprintnumbers,superscriptaddress,nofootinbib,amsmath,longbibliography,floatfix]{revtex4-1}

\usepackage{graphicx}% Include figure files 
\usepackage{dcolumn}% Align table columns on decimal point
\usepackage{bm}% bold math
\usepackage[dvipsnames]{xcolor}
\usepackage{multirow}
\usepackage{hyperref}% add hypertext capabilities
\hypersetup{
  colorlinks=true,
  citecolor=blue,
  linkcolor=blue,
  urlcolor=blue
}
\begin{document}

%\preprint{APS/123-QED}

\title{Chained Quantile Morphing with Normalizing Flows}% Force line breaks with \\

\author{Samuel Bright-Thonney}
\email{skb93@cornell.edu}
\affiliation{Physics Department, Cornell University, 109 Clark Hall, Ithaca, New York 14853, USA}

\author{Philip Harris}
\email{pcharris@mit.edu}
\affiliation{Physics Department, Massachusetts Institute of Technology, Cambridge, MA 02139, USA}
\affiliation{The NSF AI Institute for Artificial Intelligence and Fundamental Interactions, USA}

\author{Patrick McCormack}
\email{wmccorma@mit.edu}
\affiliation{Physics Department, Massachusetts Institute of Technology, Cambridge, MA 02139, USA}
\affiliation{The NSF AI Institute for Artificial Intelligence and Fundamental Interactions, USA}

\author{Simon Rothman}
\email{srothman@mit.edu}
\affiliation{Physics Department, Massachusetts Institute of Technology, Cambridge, MA 02139, USA}
\affiliation{The NSF AI Institute for Artificial Intelligence and Fundamental Interactions, USA}

\date{\today}% It is always \today, today,
             %  but any date may be explicitly specified

\begin{abstract}
Accounting for inaccuracies in Monte Carlo simulations is a crucial step in any high energy physics analysis. It becomes especially important when training machine learning models, which can amplify simulation inaccuracies and introduce large discrepancies and systematic uncertainties when the model is applied to data. In this paper, we introduce a method to transform simulated events to better match data using normalizing flows, a class of deep learning-based density estimation models. Our proposal uses a technique called \textit{chained quantile morphing}, which corrects a set of observables by iteratively shifting each entry according to a conditonal cumulative density function. We demonstrate the technique on a realistic particle physics dataset, and compare it to a neural network-based reweighting method. We also introduce a new contrastive learning technique to correct high dimensional particle-level inputs, which naively cannot be efficiently corrected with morphing strategies. 
\end{abstract}

%\keywords{Suggested keywords}%Use showkeys class option if keyword
                              %display desired
\maketitle

%\tableofcontents

\section{\label{sec:intro}Introduction}
Searches and measurements using Large Hadron Collider (LHC) data almost always rely on Monte Carlo (MC) simulations to develop analysis, validate tools, and frequently predict backgrounds. These simulations are widely acknowledged to be imperfect, particularly in modeling detector interactions and the non-perturbative physics of hadronization; data-driven strategies are preferred when possible. Despite the limitations of MC, many modern analyses rely heavily on machine learning (ML) to maximize their sensitivity, and these algorithms are typically trained on MC. The use of ML allows for the effective utilization of complex patterns and correlations in high-dimensional data and is thus more powerful, but also particularly susceptible to spurious, unphysical artifacts present in the simulations.

This issue has only grown in recent years, with a community-wide move towards training ML models on extremely granular, particle-level information with architectures such as \textsc{ParticleNet}~\cite{Qu:2019gqs,CMS-PAS-BTV-22-001}, \textsc{LundNet}~\cite{Dreyer:2020brq,ATLAS:2023krw}, and the Dynamic Reduction Network~\cite{Rothman:2799575}. As the reliance on finer details increases, simulations become less reliable. These inaccuracies can lead to significant discrepancies between a model's performance on MC and real experimental data. This adds additional work for physicists (deriving corrections, scale factors, etc.), introduces new uncertainties, and points to the deeper issue of training our most powerful analysis tools on flawed simulations. It is conceivable that, in the near future, ML-related systematic uncertainties coming from flaws in the simulation will be the primary limitation on the precision of Standard Model measurements or the sensitivity of new physics searches.

In this paper, we introduce a general purpose strategy to transform samples from one probability distribution to match another using a deep learning implementation of \textit{chained quantile morphing} (CQM)~\cite{spyromitros2016multi,qmorphcms}. CQM iteratively transforms a set of observables $\mathbf{x} = (x_1,\ldots,x_N)$ using the conditional cumulative distribution functions (CDFs) $F_i^\mathrm{MC}(x_i|\mathbf{x}_{1:i-1})$ and $F_i^\mathrm{Data}(x_i|\mathbf{x}_{1:i-1})$, and was first used for LHC analysis to improve the quality of photon identification in the Compact Muon Solenoid detector Ref.~\cite{qmorphcms}. The authors used \textit{discretized} approximations of the CDFs to correct MC inaccuracies, which allowed them to reduce an important systematic uncertainty. In this work, we develop a continuous and precise version of their approach using \textit{normalizing flows}~\cite{Kobyzev_2021,tabaknflows,papamakarios2021normalizing} -- a family of invertible ML models capable of learning complex conditional probability densities. While CQM can be used to transform between \textit{any} two distributions of the same dimensionality, we focus on the high energy physics context of transforming simulated Monte Carlo observables to better match experimental data.

Monte Carlo correction strategies typically fall into two categories: reweighting~\cite{Cranmer:2015bka,PhysRevD.101.091901,PhysRevD.103.036001,PhysRevD.102.076004,PhysRevD.105.076015,Andreassen:2020nkr} and morphing~\cite{klein2022flows,Golling:2023mqx,Golling:2022nkl,Raine:2022hht,Sengupta:2023xqy,Algren:2023qnb}. CQM is a morphing strategy, meaning we \textit{correct} the values of observables $\mathbf{x}_\mathrm{MC} \to \mathbf{x}_\mathrm{MC}^\mathrm{corr}$ in a way that results in the overall distribution agreeing better with data. Reweighting methods learn a per-event \textit{weight} that improves data/MC agreement without explicitly altering any observables. Both are effective strategies and have been extensively studied. However, the underlying motivation for each of these methodologies is different due to the nature of how they correct MC. Morphing methods will shift full distributions, thereby breaking relations of parameters within the simulation. Such a correction is applicable when considering a recalibration of a detector readout where the full distribution is shifted to correct an mis-modeled relation between a generated effect and a reconstructed effect. An example from LHC physics could be photon energy corrections, whereby one shifts the mis-reconstructed energy spectrum to match the generated spectrum.  Reweighting strategies are often used when there is a need to preserve invariant quantities such as particle mass. An example of such a strategy is reweighting simulated top quark momentum spectra to match the observed data distribution. 

While both reweighting and morphing have been shown to be effective, when distributions differ by large amounts, reweighting strategies can lead to large uncertainties in their prediction due to the presence of high weights. As a result, morphing may be more effective, particularly when there are significant differences between data and MC\footnote{One clear example is a discrepancy in the tails of data and MC distributions, where an event weight cannot make up for lack of events}. Morphing also produces a new dataset which may be easier to use in downstream applications such as training ML models. 

In the following paper, we develop and demonstrate the effective use of flow-based CQM. We discuss the details of the flow-based implementation of CQM in Sec.~\ref{sec:methods} and contrast it with existing approaches. In Sec.~\ref{sec:expts} we present results using CQM to morph between a pair of toy 2D distributions and a pair of realistic simulated particle physics datasets. Sec.~\ref{sec:contrastive} explores the possibility of applying CQM in very high-dimensional (e.g.\ particle-level) contexts by embedding the physically relevant information into a low-dimensional space using contrastive learning. Finally, in Sec.~\ref{sec:sideband} we demonstrate that CQM is insensitive to small levels of signal contamination and can be trained in a control region and interpolated accurately into a blinded signal region.

\section{\label{sec:methods}Methodology}
A normalizing flow is a density estimation model designed to learn a map $f : \mathbb{R}^d \to \mathbb{R}^d$ between an unknown training data distribution $X \sim p_D$ and a known base distribution $Z \sim p_B$ of the same dimension\footnote{See Ref.~\cite{Kobyzev_2021} for a review of modern methods}. The base distribution is typically taken to be a multidimensional standard normal $\mathcal{N}(\mathbf{0},\mathbf{1})$, and the function $f$ is constructed from a composition of invertible maps $f(\mathbf{x}) = f_N \circ f_{N-1} \circ \cdots f_1(\mathbf{x})$ with tractable Jacobian determinants. The invertible structure enables sampling the unknown distribution by sampling the base distribution, and the change of variables formula $p_D(\mathbf{x}) = p_B(f(\mathbf{x}))\left| \det \frac{\mathrm{d}f}{\mathrm{d}\mathbf{x}}\right|$ enables density estimation. Flows are trained with a log likelihood loss, which for a composite transformation, takes the form
\begin{equation*}
    \log p_D(\mathbf{x}) = \log p_B(f(\mathbf{z})) + \sum_{i=1}^{N} \log \left| \det \frac{\mathrm{d}f_i}{\mathrm{d}\mathbf{z}_{i-1}}\right|
\end{equation*}
where $\mathbf{x} = \mathbf{z}_0$, $\mathbf{z} = \mathbf{z}_N$, and $\mathbf{z}_i = f_i(\mathbf{z}_{i-1})$. The transformations $f_i$ are typically drawn from a parametrized family of functions $f_\phi$, with parameters $\phi$ computed by neural networks.

\subsection{\label{sec:nf-morph}Conditional Flows \& Quantile Morphing}
Flow models can easily be modified to fit \textit{conditional} distributions $p(\mathbf{x}|\mathbf{y})$ by allowing the parameters of the flow transformations $f_{i,\phi_i}$ to depend on the conditioning variables, i.e.\ $\phi_i = \phi_i(\mathbf{x},\mathbf{y})$. A multidimensional joint density estimation task can be decomposed into a series of one-dimensional tasks using the probability chain rule
\begin{equation}
    \label{eq:chainrule}
    p(\mathbf{x}) = p(x_1)p(x_2|x_1)\cdots p(x_d|x_1,\ldots,x_{d-1}),
\end{equation}
with each term modeled by a normalizing flow. While modern flow architectures can perform the joint estimation, this decomposition enables the analysis of (conditional) cumulative distribution functions (CDFs) for each dimension of $\mathbf{x}$ and is essential to the quantile morphing technique described in this paper.

Quantile morphing is a method to correct samples from a reference distribution $p_\mathrm{MC}$ to match those of a target distribution $p_D$\footnote{The subscripts MC and D are used here and throughout the paper as shorthand for Monte Carlo and data, respectively.} by applying a CDF transformation that matches their quantiles. In 1D, the transformation $x \mapsto y = F_D^{-1}(F_\mathrm{MC}(x))$ maps $x \sim p_\mathrm{MC}$ to $y \sim p_D$, where $F_\mathrm{MC}$, $F_D$ are the CDFs for $p_\mathrm{MC}$, $p_D$. This is an exact transformation from $p_\mathrm{MC}$ to $p_D$, and guarantees optimal transport for each shifted sample.

In higher dimensions, CDFs are not uniquely defined since rotations across dimensions are permissible, and the basic quantile morphing strategy breaks down. Fortunately, we can reconcile this ambiguity to transform $p_\mathrm{MC}$ into $p_D$ by breaking the problem into a series of 1D transformations following Eq.~\ref{eq:chainrule} via an iterative procedure called \textit{chained quantile morphing} (CQM) \cite{qmorphcms,spyromitros2016multi}. At each step, a dimension $x_i \in \mathbf{x} \sim p_D$ is morphed to $y_i \sim p_D(y_i)$ using the \textit{conditional} quantile functions $F_\mathrm{MC}(x_i|y_1,y_2,\ldots,y_{i-1})$ and $F_D(y_i|y_1,y_2,\ldots,y_{i-1})$. Starting from the first dimension $x_1$, chained quantile morphing proceeds as follows:
\begin{enumerate}
    \item Transform $x_1$ to $y_1 = F_D^{-1}(F_\mathrm{MC}(x_1))$ using CDFs $F_\mathrm{MC}$ and $F_D$.
    \item Transform $x_2$ to $y_2 = F_D^{-1}(F_\mathrm{MC}(x_2|y_1)|y_1)$ using conditional CDFs $F_\mathrm{MC}(\cdot|y_1)$ and $F_D(\cdot|y_1)$ and the corrected value $y_1$ from step (1).
    \item Continue as in (2) for $i = 3,\ldots,d$ with $y_i = F_D^{-1}(F_\mathrm{MC}(x_i|\mathbf{y}_{1:i-1})|\mathbf{y}_{1:i-1})$ and the previously corrected dimensions $\mathbf{y}_{1:i-1}$.
\end{enumerate}

If all (conditional) CDFs are known analytically, the CQM procedure guarantees an exact transformation of samples $\mathbf{x} \sim p_\mathrm{MC}$ to $\mathbf{y} \sim p_D$. In any real world application, however, the reference and target datasets will come from complex and/or high-dimensional distributions with intractable densities and CDFs. In these cases, the CDFs can only be approximated, typically by training machine learning (ML) algorithms to perform conditional quantile regression.

CQM was first introduced for LHC physics for photons within the CMS detector in Ref.~\cite{qmorphcms}, where it was used to correct mis-modeled MC to better match experimental data through the use of a well defined control region within data. The authors constructed a discretized approximation of each CDF by training boosted decision trees (BDTs) to learn a fixed array of conditional quantiles. This was effective for the analysis but difficult to scale efficiently due to the fixed quantiles and the large number of BDT trainings required. In this work, we propose a streamlined approach to CQM with normalizing flows (NFs). Flows can learn the full conditional density (and thus CDF) at each step of CQM, enabling a more exact version of the transformation.

\subsection{CQM with Normalizing Flows \label{sec:flowcqm}}
\begin{figure}[t]
\includegraphics[width=\linewidth]{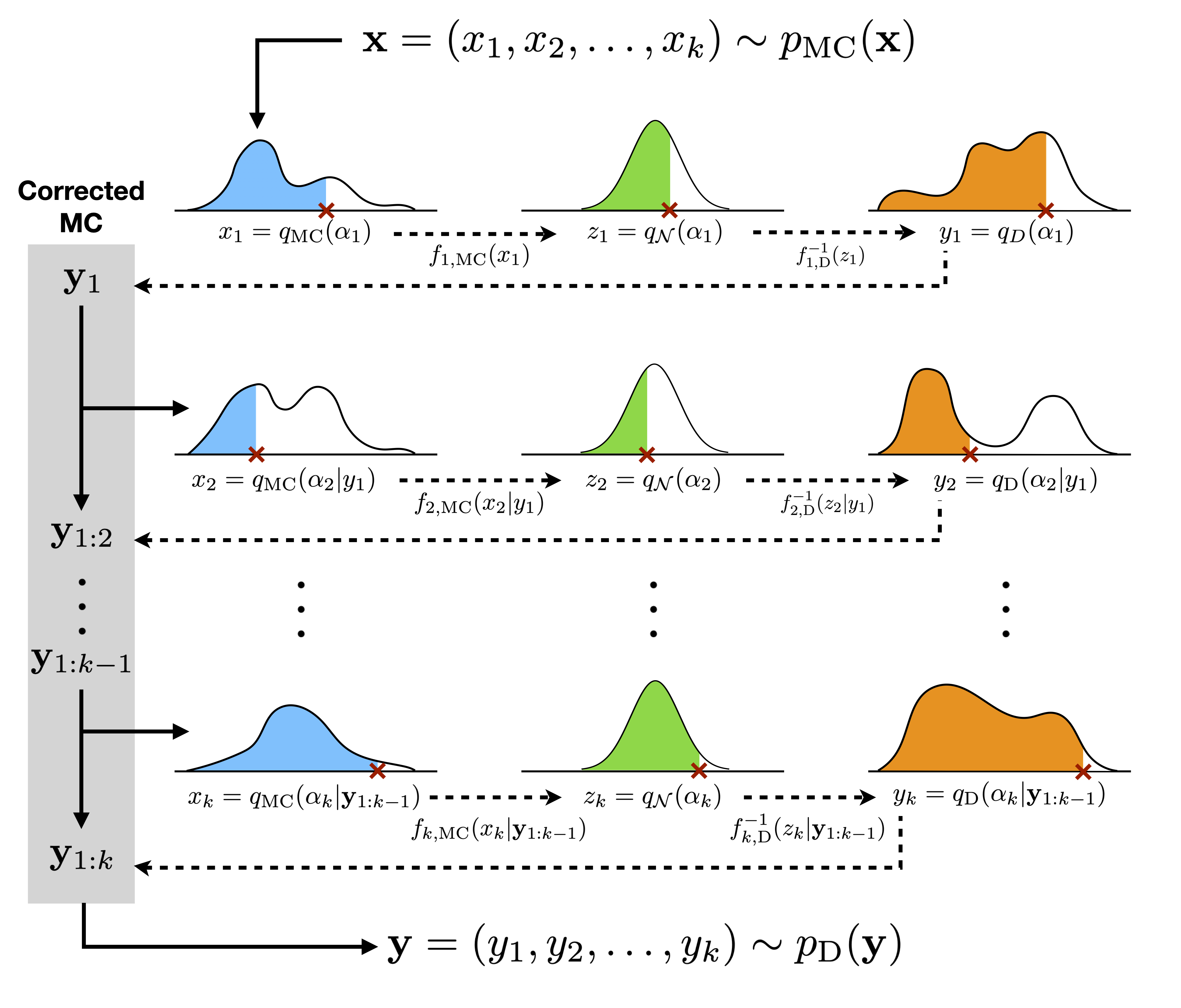}
\caption{\label{fig:morph}A schematic demonstrating how chained quantile morphing transforms samples from one $k$-dimensional PDF $p_\mathrm{MC}(x_{1},\ldots,x_{k})$ to another $p_D(x_1,\ldots,x_k)$. The shaded regions in blue, green, and orange denote the conditional quantiles of variables $x_i$, $z_i$, $y_i$, and the red X's mark their values. CQM matches the conditional quantiles of the original distribution $p_\mathrm{MC}$ to those of the target distribution $p_\mathrm{D}$ by mapping $x_i = q_\mathrm{MC}(\alpha_i|\mathbf{y}_{1:i-1}) \mapsto y_i = q_\mathrm{D}(\alpha_i|\mathbf{y}_{1:i-1})$, preserving the $p$-value $\alpha_i$.}
\end{figure}
Figure \ref{fig:morph} shows a schematic of the CQM correction procedure implemented with normalizing flows. Given datasets $\{\mathbf{x}_j\}_\mathrm{MC}$ (``Monte Carlo") and $\{\mathbf{y}_j\}_D$ (``data"), the components $x_i \in \mathbf{x} \sim p_\mathrm{MC}$ are iteratively corrected to match the distributions of $y_i \in \mathbf{y} \sim p_D$ using the conditional probability decomposition shown in Eq.~\ref{eq:chainrule}. At step $i$, flows $f_{i,S}$ and $f_{i,D}$ are trained to fit the conditional distributions $p_\mathrm{MC}(x_i|\mathbf{x}_{1:i-1}^\mathrm{corr})$ and $p_D(y_i|\mathbf{y}_{1:i-1})$, respectively, where $\mathbf{x}_{1:i-1}^\mathrm{corr}$ are the dimensions of $\mathbf{x}$ corrected in previous steps. Individual points $x_{i,k}$ are then corrected by the transformation 
\begin{equation}
    x_{i,k}^\mathrm{corr} = f_{i,D}^{-1}(f_{i,\mathrm{MC}}(x_{i,k} | \mathbf{x}_{1:i-1,k}^\mathrm{corr})|\mathbf{x}_{1:i-1,k}^\mathrm{corr})
\end{equation}

This procedure leverages the conditional \textit{flow} in lieu of the conditional quantile function, but the quantile morphing operation is fundamentally the same as described in Sec.~\ref{sec:nf-morph}. A given data point $x_{i,k} \in \mathbf{x}_k$ will correspond to some conditional quantile of the distribution $p_\mathrm{MC}(x_i|\mathbf{x}^\mathrm{corr}_{1:i-1})$, and a faithfully trained conditional flow $f_{i,S}$ will map it to the same quantile $z_{i,k}$ of a standard normal distribution $\mathcal{N}(0,1)$. The inverse flow $f_{i,D}$ will then map $z_{i,k}$ to the corresponding quantile of $p_D(y_i|\mathbf{x}_{1:i-1}^\mathrm{corr})$. After the full chain of corrections is applied, the samples $\{\mathbf{x}_j^\mathrm{corr}\}$ will follow the target distribution $p_D$.

\subsection{Flow Implementation \label{sec:implementation}}
We parameterize our flow transformations $f_\phi$ using piecewise rational quadratic splines~\cite{durkan2019neural} implemented with \texttt{PyTorch}~\cite{paszke2019pytorch} in the \texttt{nflows} package~\cite{nflows}. The splines are defined on the interval $[-3.2,3.2]$, and all input data are min-max scaled to the range $[-3,3]$ to capture the tails. All flows are trained with the \texttt{AdamW} optimizer~\cite{kingma2017adam} on a cosine-annealed learning rate schedule~\cite{loshchilov2017sgdr}. The spline parameters for each flow transformation are determined from the conditioning inputs using a neural network\footnote{When there are no conditioning inputs (i.e.\ the first flow of the chain), the network is simply passed zeros.}. 

Although we implement CQM with a sequence of \textit{distinct} flows, we note that it is possible to use an autoregressive architecture such as \textsc{MADE}~\cite{germain2015made} to train a single flow that simultaneously learns each conditional density of Eq.~\ref{eq:chainrule}. While this streamlines the process, we found that it was generally quite difficult to achieve simultaneous high-quality fits to all inputs using this approach.

\subsection{Comparison with Existing Strategies}
Our approach to chained quantile morphing builds on the previous approach with BDTs~\cite{qmorphcms}, and extends the scope of this effort through the use of conditional normalizing flows. Additionally, a large variety of machine learning-based morphing and reweighting strategies have emerged in recent years \cite{Cranmer:2015bka,PhysRevD.101.091901,PhysRevD.103.036001,PhysRevD.102.076004,PhysRevD.105.076015,Andreassen:2020nkr,klein2022flows,Golling:2023mqx,Golling:2022nkl,Raine:2022hht,Sengupta:2023xqy,Algren:2023qnb}, including several proposals using normalizing flows~\cite{klein2022flows,Golling:2023mqx,Golling:2022nkl,Raine:2022hht,Sengupta:2023xqy,Algren:2023qnb}\footnote{For an overview of newer methods, see Ref.~\cite{Golling:2023yjq}}. Reweighting methods aim to correct MC to match data by training a supervised learning algorithm to differentiate the two datasets, then reweighting the simulated dataset based on the discriminator output~\cite{hastie2009elements,sugiyama_suzuki_kanamori_2012}. Reweighting and quantile-morphing methods have strengths for different types of corrections, which we briefly investigate in App.~\ref{ap:reweight}. Broadly speaking, we find reweighting to be more effective in correcting for theoretical physics generator mis-modeling and quantile morphing more effective for fixing simulated detector mis-modeling. 

CQM is most similar to the ``flows for flows" (FFF) paradigm, proposed in Ref.~\cite{klein2022flows} and rephrased several times in~\cite{Golling:2023mqx,Golling:2022nkl,Golling:2023yjq,Sengupta:2023xqy}, but differs in two key ways. First, FFF transforms $\mathbf{x} \sim p_\mathrm{MC}$ to $\mathbf{y} \sim p_D$ all at once using \textit{joint} density estimation, rather than transforming the variables $x_i$ iteratively as in CQM. Secondly, FFF operates by using the learned density $p_D$ as a target for training a flow transformation that maps $p_\mathrm{MC}$ directly to $p_D$. CQM trains separate flows for $p_\mathrm{MC}$ and $p_D$, and transforms events $\mathbf{x} \sim p_\mathrm{MC}$ via the shared Gaussian latent space.
%PH: I don't think we need to overly discuss FFF, I cut it down
%Each approach has benefits and drawbacks, and it is beyond the scope of this paper to perform a detailed comparison. While CQM requires more training steps than FFF (one per variable), it allows finer control over the quality of each fit. If a particular variable is difficult to morph accurately due to intricate correlations or an insufficiently expressive flow architecture, it can be re-trained without impacting previously corrected variables. 
With FFF, any unsatisfactory fit would require a re-training of the full joint density, whereas CQM requires a single variable re-training. In cases where a very high quality fit is needed -- minimizing systematic uncertanties, for example -- this level of fine-tuned control may be advantageous.CQM additionally imposes a notion of \textit{local} transformations, as each 1D quantile map will typically shift a variable by a relatively small distance. Other methods, such as FFF, can only target locality by introducing an ad-hoc modification to the loss function during training~\cite{Golling:2023mqx}.

\section{Experiments\label{sec:expts}}
Chained quantile morphing is a general-purpose technique for mapping between any two densities, and can applied in a wide variety of contexts. In this section, we explore its performance in two use cases: a toy example mapping between two-dimensional datasets and a high-energy physics application using high-level observables from simulated proton-proton collisions. In the latter example we also explore how the transformed samples from CQM can be used in downstream applications, namely training neural networks.

\subsection{Toy 2D Dataset \label{sec:toy2d}}
\begin{figure}[t]
    \centering
    \includegraphics[width=0.45\linewidth]{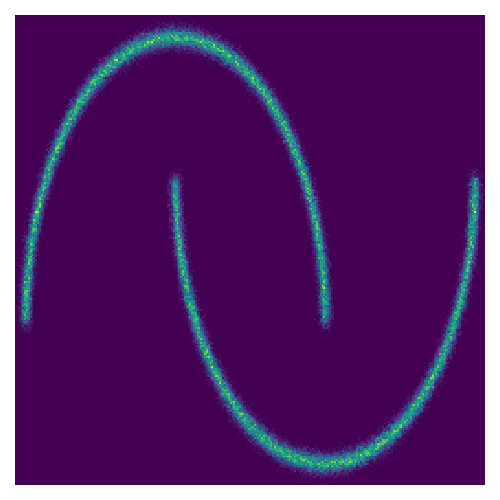}
    \includegraphics[width=0.45\linewidth]{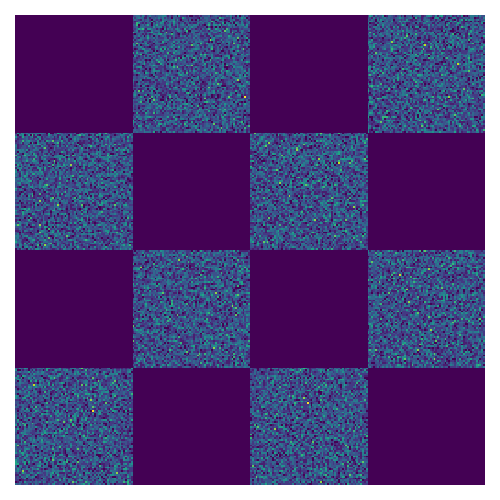}
    \\
    \includegraphics[width=0.45\linewidth]{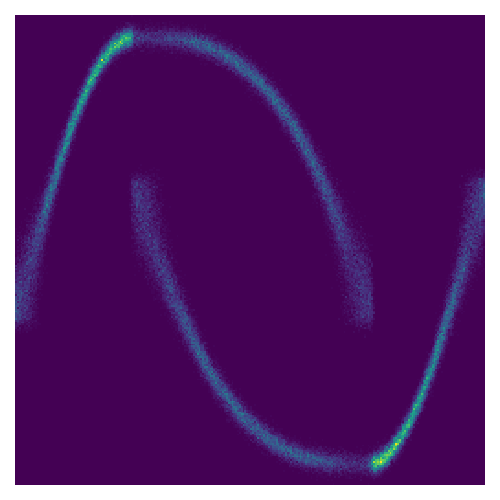}
    \includegraphics[width=0.45\linewidth]{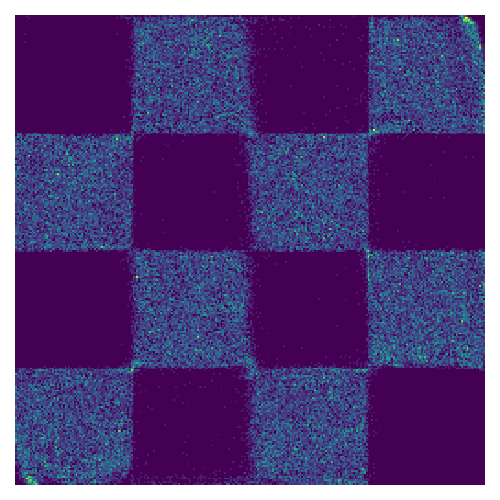}
    \caption{\label{fig:toy}Using CQM to morph between a pair of two-dimensional datasets (top row). The bottom row shows the transformed distribution after morphing the $x$-axis (left) then the $y$-axis (right).}
\end{figure}

As a simple example, we first demonstrate CQM on a pair of two-dimensional datasets shown in Fig.~\ref{fig:toy}. The top row shows the reference (half moons) and target (checkerboard) densities, and the bottom row shows each step of the CQM transformation. In the first step the $x$ distributions are matched, notably thinning the density in the parts of each half moon that overlap in the $y$ direction. In the second step the $y$ distributions are matched, faithfully reproducing the sharp boundaries and alternating conditional pattern of the target density. 

This example demonstrates the flexibility of flow-based CQM, and highlights the advantage of using a density estimation approach that can readily capture complex single-variable distributions and strong conditional dependencies between inputs. Most realistic applications will have weaker correlations between input dimensions and less exotic marginals.

\subsection{LHC Olympics Datasets \label{sec:lhco}}
We now demonstrate how CQM can be applied in high-energy physics using the LHC Olympics (LHCO) datasets \cite{lhco}. We focus on the LHCO ``R\&D" \cite{lhco_rd} and ``Black Box 1" (BB1) \cite{lhco_official} datasets, and train CQM to map between them. The R\&D dataset contains one million Standard Model quantum chromodynamics (QCD) background dijet events, and 100,000 signal dijet events featuring a heavy resonance decay $Z^\prime \to XY$ ($X\to q\bar{q}$, $Y\to q\bar{q}$) with $m_Z = 3.5$ TeV \cite{Kim:2019rhy}. The BB1 dataset contains a total of one million events, 834 of which are signal events with the same topology as the R\&D signal but with different resonance masses. Both datasets are generated with \texttt{Pythia 8.219} \cite{Bierlich:2022pfr,Sjostrand:2006za} and \texttt{Delphes 3.4.1} \cite{deFavereau:2013fsa,Mertens:2015kba}, but with different parameter configurations that alter the distributions of relevant high-level jet observables.

We cluster the particles in each event into jets using the anti-$k_T$ algorithm \cite{Cacciari:2008gp} with radius $R = 0.8$, and define jet 1 (2) to be the heaviest (second heaviest) jet in the event. For our high-level observables, we compute $\rho = m_J/p_{T,J}$ and the $N$-subjettiness ratios $\tau_{21}$, $\tau_{32}$, and $\tau_{43}$~\cite{Thaler_2011} as a measure of their 2, 3, or 4-pronged substructure.

Using only QCD background events from each dataset (i.e.\ removing the signal events from BB1), we train the CQM procedure to map between them in the eight-dimensional space constructed from $(\rho, \tau_{21}, \tau_{32}, \tau_{43})$ of each jet.\footnote{Quantile morphing can be performed in either direction due to the invertibility of normalizing flows, i.e. R\&D $\to$ BB1 or BB1 $\to$ R\&D} We use 500,000 R\&D and 500,000 BB1 events to train CQM, and evaluate its performance with the remaining events.

\begin{figure*}
\includegraphics[width=\textwidth]{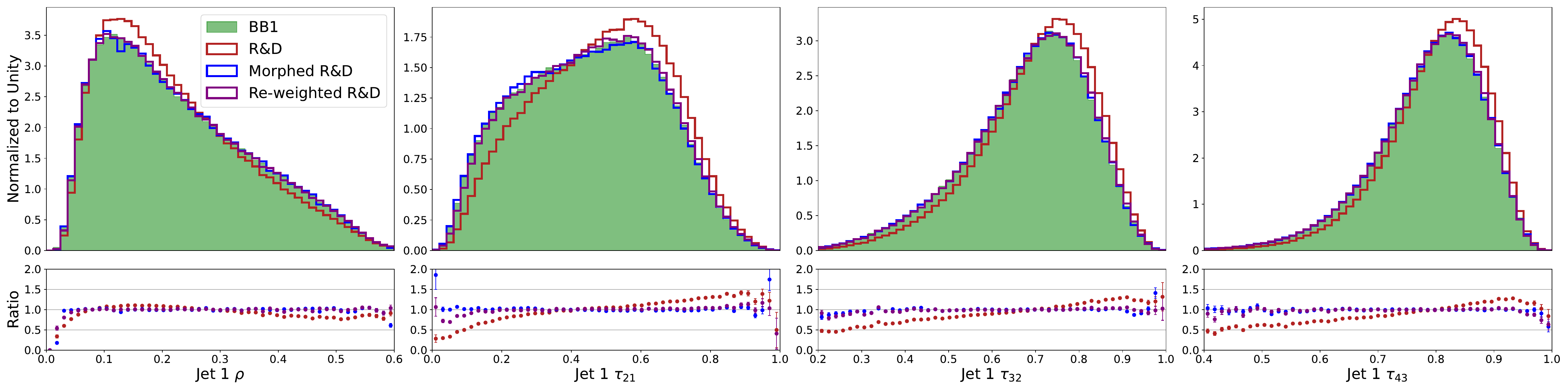}
\includegraphics[width=\textwidth]{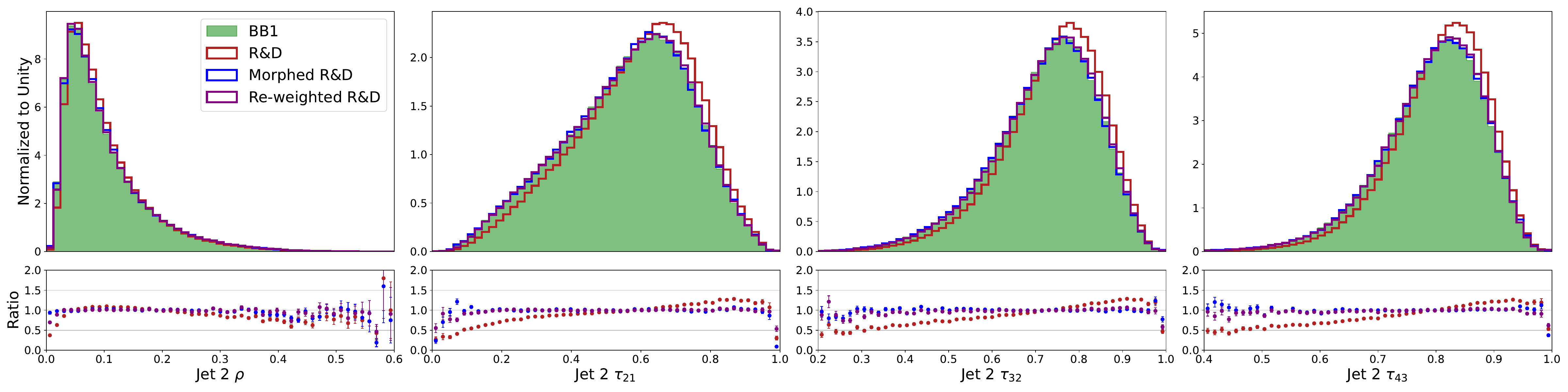}
\caption{\label{fig:1dcorrections}Distributions of $\rho, \tau_{21}, \tau_{32}, \tau_{43}$ (left to right) of the leading (top row) and sub-leading (bottom row) jets in the BB1 (green), R\&D (red), and morphed R\&D (blue) datasets. The morphed R\&D data is a very close match to BB1, as shown by the ratio plots in the bottom panes. We also show the results of a neural network-based reweighting of R\&D data, which yields similarly good agreement.}
\end{figure*}

Fig.~\ref{fig:1dcorrections} shows the distributions of each input variable from the test sets before and after morphing R\&D to match BB1. The morphed distributions are a very good visual match to the target distributions, and the ratio plots indicate good agreement across nearly the entire range of each input. Performance begins to break down in the extreme high and low tails, but this is to be expected due to statistical fluctuations and limited training samples available in these sparse regions. We also show results of a standard neural network-based reweighting scheme, implemented with a six-layer fully connected network with \texttt{ReLU} activations, a Sigmoid output, and 10\% dropout to prevent overfitting. The network is trained to discriminate R\&D events from BB1 events using a binary cross entropy loss, and the score $s(\mathbf{x})$ is converted into an event weight $w(\mathbf{x}) = s(\mathbf{x})/(1-s(\mathbf{x}))$~\cite{hastie2009elements,sugiyama_suzuki_kanamori_2012}. As expected, the reweighting performs on-par with CQM. Unlike CQM, however, it does not produce a set of transformed samples that can be used for downstream tasks.

To better quantify the quality of the morphed distribution, we train neural networks to distinguish BB1 samples from R\&D samples before and after corrections are applied. In Table \ref{table:aucs} we report the area under the curve (AUC) metrics, each averaged over an ensemble of 10 trainings. The AUCs drop to 0.51 after morphing in either direction, indicating that CQM produces samples that are virtually indistinguishable from the target distribution. We also evaluate the reweighting by computing weighted AUCs for the nominal training (i.e.\ un-morphed samples), and find that it also succeeds. 

\begin{table}
    \centering
    \begin{tabular}{|c|c|}
        \hline
        Comparison & AUC \\
        \hline\hline
        BB1 vs.\ R\&D & $0.6257 \pm 0.0002$ \\
        Morphed BB1 vs.\ R\&D & $0.5093 \pm 0.0004$ \\
        BB1 vs.\ Morphed R\&D & $0.5106 \pm 0.0004$ \\
        BB1 vs.\ Reweighted R\&D & $0.5158 \pm 0.0001$ \\
        \hline
    \end{tabular}
    \caption{\label{table:aucs}Area under the curve (AUC) metrics for neural network discriminators trained to distinguish pairs of datasets (BB1 and R\&D) before and after applying chained quantile morphing to make one match the other. The reported AUCs are computed from the average of 10 separate trainings. We also evaluate re-weighting with a weighted AUC using the BB1 vs.\ R\&D training.}
\end{table}

\subsection{Morphed Distributions for Classification Tasks \label{sec:sigvsbkg}}
As demonstrated in the previous section, chained quantile morphing can transform samples from the reference distribution into samples indistinguishable from the target distribution. This is a powerful tool for correcting simulation inaccuracies in high-level observables, and can find immediate use in reducing systematic uncertainties and easing cut-based analysis workflows. Beyond these standalone applications, however, CQM-transformed simulation can be used in more complex downstream tasks such as training neural networks to separate signal and background.

We demonstrate this use case using the same LHCO samples analyzed in Sec.~\ref{sec:lhco}. To emulate an LHC analysis, we treat the R\&D background dataset as ``background simulation", the R\&D signal as ``signal simulation", and the BB1 background as ``data". We use CQM to construct a morphed R\&D background dataset to match the BB1 (``data") background, and train separate neural network classifiers to distinguish the morphed and un-morphed samples from the R\&D signal\footnote{See Sec.~\ref{sec:sideband} for a similar study including signal contamination in the target ``data" sample.}. We evaluate each classifier on the four background samples: R\&D, BB1, morphed R\&D, and morphed BB1. This provides insight into (a) how classifier performance changes when moving from ``simulation" to ``data", and (b) if training/testing on morphed samples replicates the performance on the target datasets they are meant to match.

\begin{figure}
    \centering
    \includegraphics[width=\linewidth]{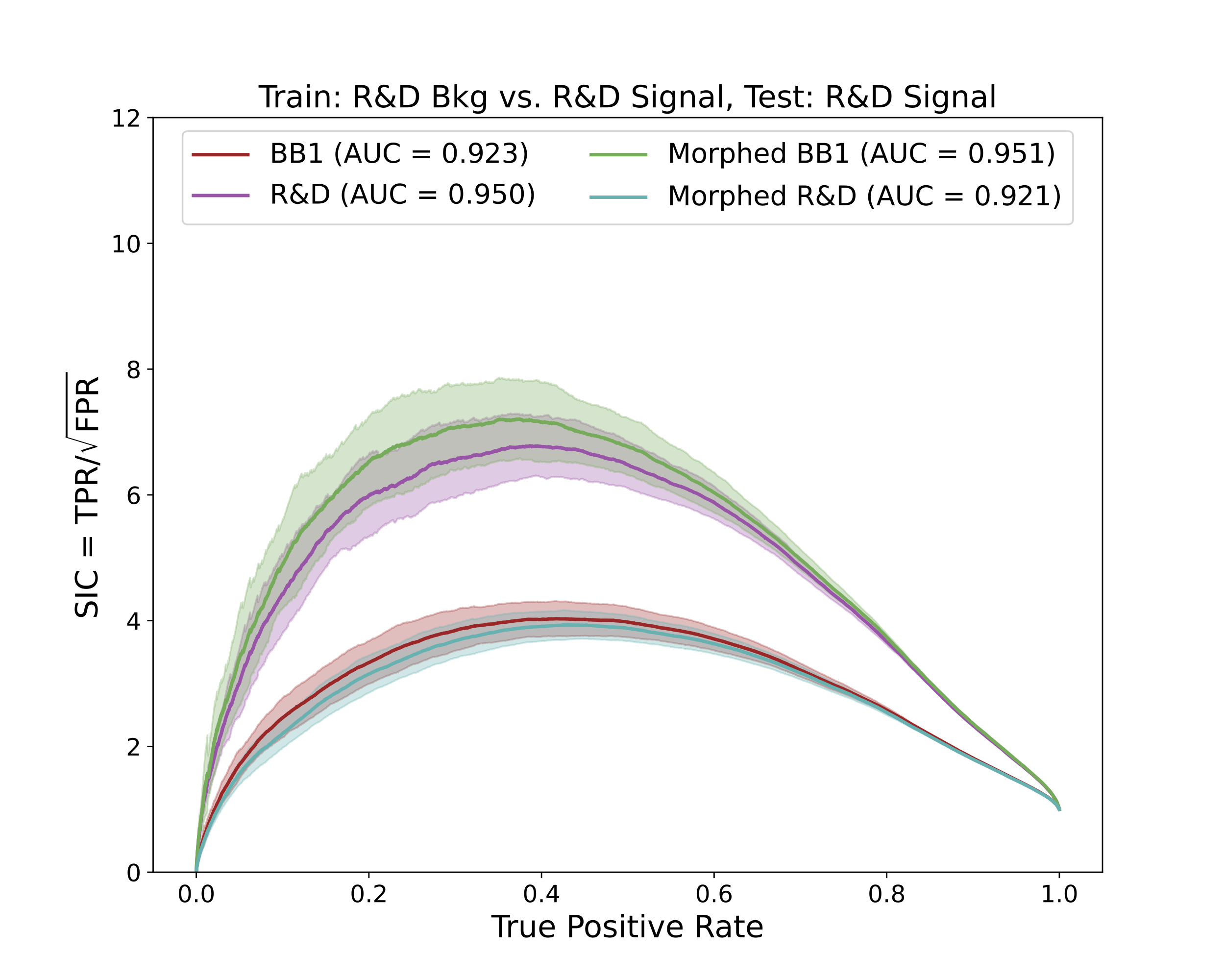}
    \includegraphics[width=\linewidth]{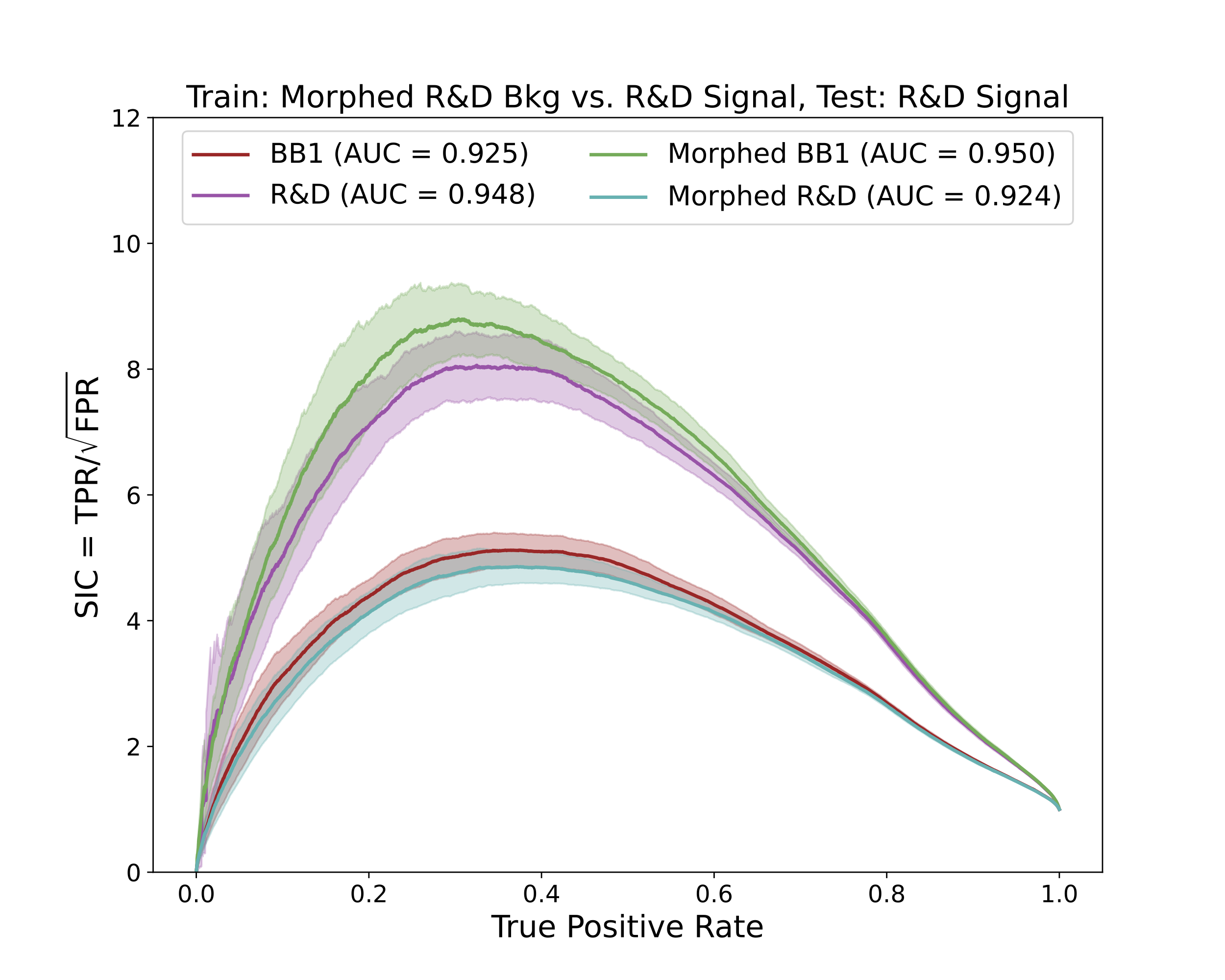}
    \caption{\label{fig:sics} Significance improvement characteristic (SIC) curves for neural networks trained to distinguish R\&D background from R\&D signal (top), and morphed R\&D background from R\&D signal (bottom). Each curve is computed by evaluating the classifier on a mixture of R\&D signal events with background-only BB1 (red), R\&D background (purple), morphed background-only BB1 (green), and morphed R\&D background (teal).}
\end{figure}

In Fig.~\ref{fig:sics}, we plot significance improvement characteristic (SIC) curves for each training/evaluation, where $\mathrm{SIC} = \frac{\mathrm{TPR}}{\sqrt{\mathrm{FPR}}}$. The upper plot shows results for trainings using the original R\&D background, and colored lines correspond evaluations on the various background datasets. The lower plot shows the same thing, but for trainings using the \textit{morphed} R\&D background. For a fixed evaluation dataset, the performance is not significantly different between trainings, but is not necessarily expected to be. Crucially, the performance is extremely consistent between BB1 and morphed R\&D, and between R\&D and morphed BB1. In the LHC physics context, CQM would allow us to predict a classifier's performance on data (BB1) by morphing MC simulation (R\&D) to match data. Curiously, the classification performance is uniformly better on the R\&D/morphed BB1 samples than the BB1/morphed R\&D samples. This simply indicates that the R\&D background is more separable from signal than the BB1 background \footnote{This is because the $\tau_{21}$ is skewed lower in the BB1 sample, as shown in Fig.~\ref{fig:1dcorrections}, and both jets in the $Z^\prime \to XY$ signal are expected to have two prongs (i.e.\ low $\tau_{21}$)}. 

Although this example is relatively simplistic (a supervised search for a resonant signal), it demonstrates CQM's potential for transforming imperfect simulations into a readily usable emulation of data. This enables a reliable estimate of an analysis' performance on real data and relies solely on having a suitable reference dataset, such as a signal free control region with similar background kinematics to the signal region. As we will discuss in Sec.~\ref{sec:sideband}, the technique is robust against modest signal contamination in the reference dataset and can be configured to interpolate from a kinematic sideband into a signal region by conditioning on the relevant variables defining the sideband (e.g.\ dijet mass $M_{jj}$).

\section{High-Dimensional CQM using Contrastive Learning\label{sec:contrastive}}
\subsection{Handling large input spaces\label{sec:contrastive-into}}
Chained quantile morphing is ideally suited for correcting $\mathcal{O}(10)$-dimensional input spaces constructed from high-level physics observables. The addition of more variables complicates the construction of the conditional probability distributions, making it harder to learn. An input space that is  $\mathcal{O}(10)$-dimensional is already very useful for many LHC physics applications that rely on high-level observables such as precision measurements. However, a generalized high-dimensional approach does not naturally result from this approach. Many of the cutting-edge jet taggers deployed at ATLAS and CMS use low-level particle- or detector-level inputs and therefore have $\mathcal{O}(1000)$-dimensional input spaces. Applying CQM to these inputs is infeasible due to the number of trainings required and the fundamental ambiguity of ``correcting" a variable-length unordered collection of particle kinematics.

Despite CQM being ineffective here, resolving data/MC discrepancies in low-level taggers is of particular interest since granular event information often amplifies the impact of MC mis-modeling and leads to larger systematic uncertainties. While low-level taggers perform significantly better than high-level taggers, the large mis-modeling can almost completely remove the improvement present. To overcome this, we rely on new approaches in deep metric learning to learn an intermediate low-dimensional space with compelling features that would be effective for CQM. Given such an embedding, data/MC discrepancies could be addressed by applying CQM in the \textit{feature space}, rather than on the level of individual particles. 

We propose to perform this with \textit{contrastive learning} algorithms~\cite{lecunContra1993,lecunContra2005,chen2020simple,dillon2021symmetries,Dillon_2022}, which learn through self-supervision to embed input data into a structured feature space where different classes of inputs (e.g. light quark jets vs.\ $b$ jets) are well separated. Assuming that the feature space captures the relevant classification information from the low-level inputs, an ML tagger trained on the resulting feature space should match the performance of a tagger trained directly on the low-level inputs. 

\subsection{LHCO implementation and experiments}
To demonstrate this approach, we train a contrastive feature space to separate signal and background in the LHCO R\&D dataset. As low-level inputs, we use the $(p_{T},z,\Delta\eta,\Delta\phi)$ of the 100 highest-$p_T$ constituents of the leading (heaviest) jet in each event, where $z = p_T/p_{T,\mathrm{jet}}$, $\Delta \eta = \eta - \eta_\mathrm{jet}$, and $\Delta\phi = \phi - \phi_\mathrm{jet}$. We use the ParticleNet~\cite{Qu:2019gqs} architecture as an embedding function into a four-dimensional feature space, and train it on a mixture of signal and background events using the VICReg (``Variance-Invariance-Covariance Regularization") contrastive loss~\cite{bardes2022vicreg}. VICReg learns a non-trivial representation by rewarding functions that embed ``like pairs" near to one another (invariance), while maintaining decorrelated and suitably spread out features (variance/covariance). In typical contrastive learning setups, training pairs correspond to a training input and an ``augmented" version of that input (e.g.\ an image and a randomly rotated version of the same image). In our implementation, we simply use pairs of signal and pairs of background events to train the algorithm. This method effectively constructs a mostly decorrelated feature space that captures the separation between signal and background. The construction of a largely decorrelated feature space is appealing for CQM, and we can use this feature space as an intermediate space to correct data to simulation. 

\begin{figure*}
    \centering
    \includegraphics[width=0.45\textwidth]{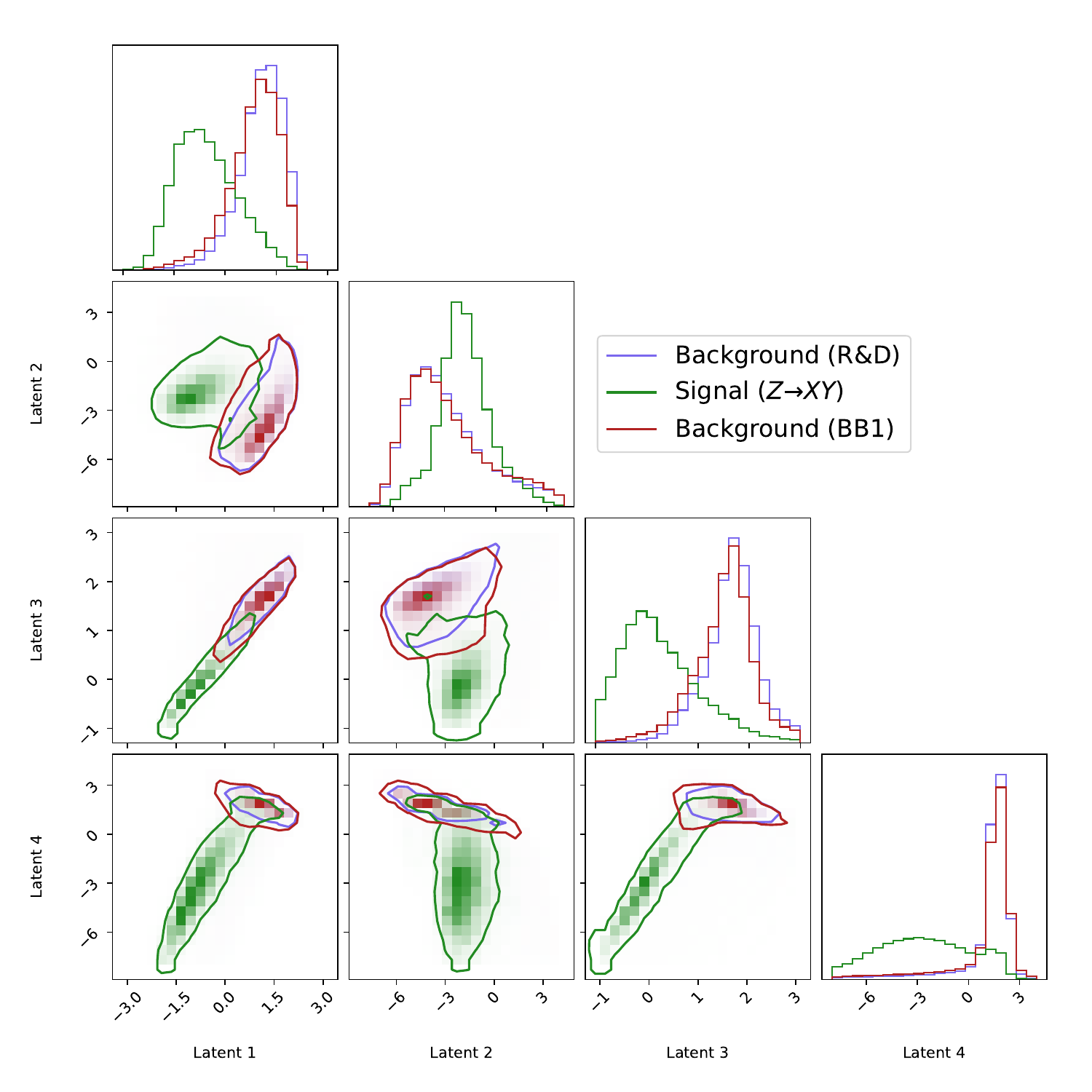}
    \includegraphics[width=0.45\textwidth]{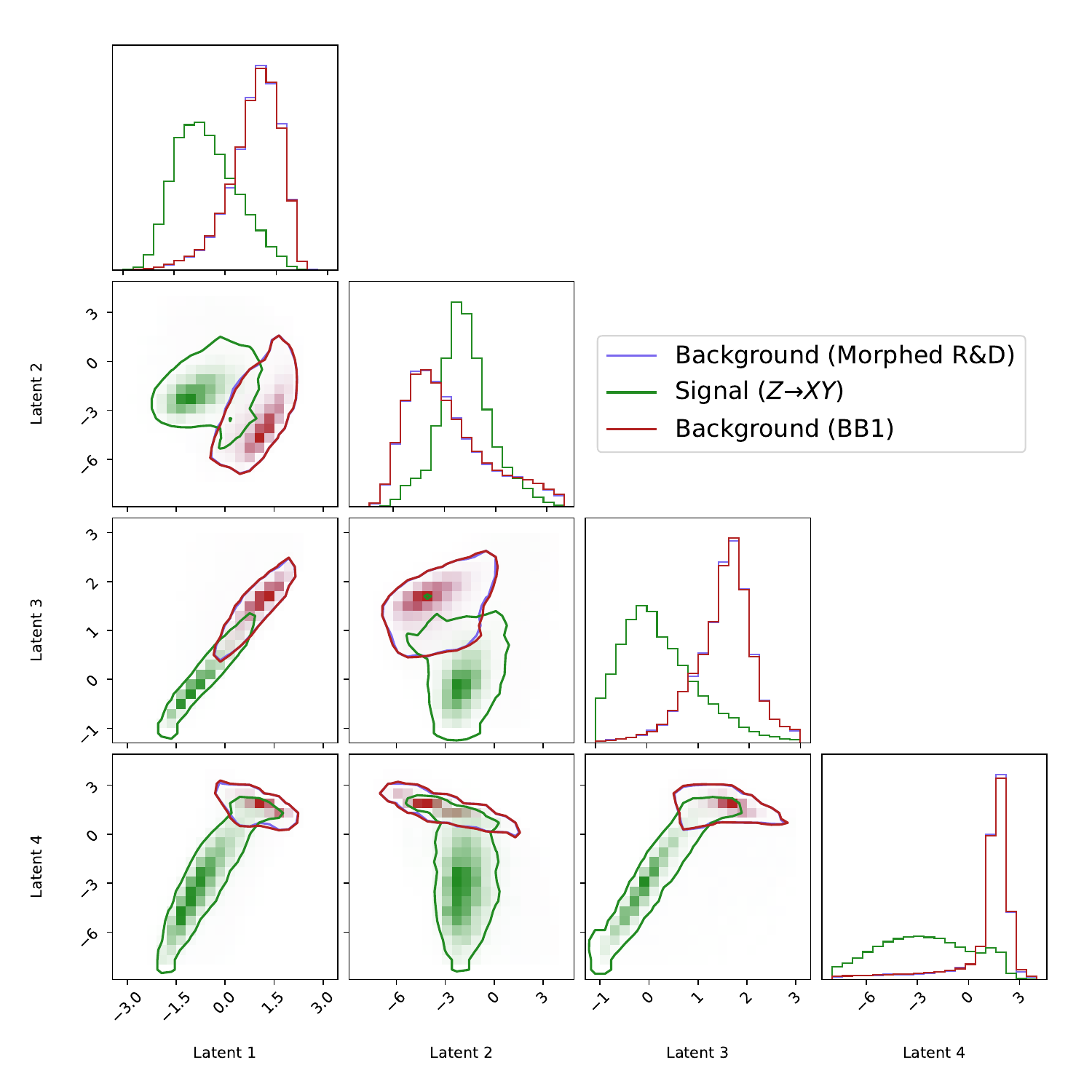}
    \caption{\label{fig:contrastive-space} (Left) The resulting latent space constructed from self-supervised VICReg training using the R\&D background sample and the $Z^\prime \to XY$ signal. The red shows embeddings of the background-only BB1 sample appears, which disagree slightly with the R\&D backgrounds. (Right) The same plot, but after applying CQM to transform the R\&D background to match the BB1 background in the latent space.}
\end{figure*}

To assess the quality of the contrastive space, we train an MLP classifier to distinguish embedded signal/background events and compare its performance with that of a ParticleNet classifier trained directly on the low-level features. We then use CQM to morph embedded R\&D background events to match embedded BB1 background events and evaluate them using the MLP classifier. Finally, we train an MLP tagger using only the high-level variables $(\rho,\tau_{21},\tau_{32},\tau_{43})$ of the leading jet to serve as a benchmark. The results are summarized in Table \ref{tab:contrastive-performance}, and the embedding spaces are shown in Fig.~\ref{fig:contrastive-space}. The tagger trained on embeddings matches the performance of the low-level tagger, and applying CQM in the contrastive space transforms the R\&D events to closely match BB1. By combining contrastive embedding and CQM, we are able to retain the advantages of using low-level inputs while simultaneously being able to correct undesired differences between two samples. In the LHC context, this is a promising new avenue for correcting data/MC discrepancies in low-level jet taggers and reducing the associated systematic uncertainties. 

\begin{table}[]
    \centering
    \begin{tabular}{|c|c|c|}
        \hline
        Classifier & Sample & AUC \\
        \hline 
        \hline 
        \multirow{2}{*}{ParticleNet Tagger} & R\&D & 0.954 \\ & BB1 & 0.931\\
        \hline
        \multirow{3}{*}{Embeddings (MLP)} & R\&D & 0.952 \\ & BB1 & 0.929 \\ & Morphed R\&D & 0.928 \\
        \hline
        \multirow{3}{*}{High-Level (MLP)} & R\&D & 0.910 \\ & BB1 & 0.872 \\ & Morphed R\&D & 0.873 \\
        \hline
    \end{tabular}
    \caption{\label{tab:contrastive-performance}A table showing the performance of the three taggers described in the text, evaluated by the AUC for separating R\&D signal from the indicated background sample. ParticleNet (top) is trained on low-level features, where as the other two are trained on contrastive embeddings (middle) and high-level jet features (bottom).}
\end{table}

\section{Signal Contamination \& Sideband Studies \label{sec:sideband}}
The studies presented thus far morph an inclusive ``MC" background sample to an inclusive ``data" sample with the assumption that the underlying physics in both samples are the same. In most realistic LHC analyses, however, MC samples are used to estimate backgrounds in regions of phase space where the data is expected to contain (in the case of measurements) or may contain (in the case of searches) additional physics unaccounted for by the simulations. It is thus unrealistic to simply morph between two inclusive datasets. Instead, we would like to use a signal-free control region to learn the morphing functions, and then interpolate or extrapolate these functions into the signal region. In this section, we present two studies addressing this question.

First, we consider the case of interpolating the morphing function trained in a control region. As in Refs.~\cite{Golling:2022nkl,Golling:2023mqx,Golling:2023yjq} and in keeping with the LHCO dataset, we consider the case of a resonant signal decaying to a pair of jets. We take the resonant mass to be $M_\mathrm{res} = 3.5$ TeV, define our signal region (SR) as $M_{jj} \in [3.3,3.7]$ TeV, and train CQM in the $M_{jj}$ sideband (SB). We again treat the R\&D dataset as ``MC" and signal-free BB1 as ``data", and transform the variables $(\rho,\tau_{21},\tau_{32},\tau_{43})$ of the leading jet. All flows are conditioned on $M_{jj}$ to allow interpolation into the signal region. In Table \ref{tab:interp-morphing} (top row), we show AUCs for a classifier trained to distinguish signal-free BB1 from morphed R\&D in the SB and SR. The AUCs indicate that they are indistinguishable in both the sideband and signal region, demonstrating that CQM remains effective when interpolating into a region of phase space unseen during training. In this case, the SR was defined by a cut on a single variable, $M_{jj}$, but this could easily be generalized to SRs defined by multiple observables.

As a second test, we consider the case where the signal is small relative to the background. If the signal is sufficiently small, it contributes so negligibly to the single variable ``data" density learned by CQM that it is unnecessary to define cuts that remove it.  To test this, we train using the R\&D background sample as ``MC" and the \textit{full} BB1 dataset as ``data" -- i.e.\ without signal events artificially removed -- with a signal contamination fraction of 0.08\%. Table \ref{tab:interp-morphing} (bottom) indicates that the morphed R\&D sample is still essentially indistinguishable from the uncontaminated BB1 background in this case, showing that a small degree of signal contamination has little to no impact on the quality of the morphed samples.  The impact of small contamination is particularly robust for CQM when no single variable dominates the discrimination and the discrimination is spread over a variety of correlated variables. Since CQM is applied one variable at a time, sensitivity to ``signal regions" is particularly reduced, making CQM robust in the regions of high separation between signal and background.

\begin{table}[t]
    \centering
    \begin{tabular}{|c|c|c|}
        \hline
        Training & Selection & AUC\\
        \hline 
        \hline 
        \multirow{2}{*}{Sideband} & SB & $0.514 \pm 0.001$  \\ & SR & $0.510 \pm 0.004$ \\
        \hline 
        Contaminated & Inclusive & $0.514 \pm 0.001$ \\
        \hline
    \end{tabular}
    \caption{\label{tab:interp-morphing}Evaluating the quality of the morphed R\&D $\to$ BB1 samples when CQM is configured to interpolate into a blinded signal region defined by an $M_{jj}$ (top), or trained using a slightly contaminated ``data" sample (bottom). As before, the AUC is computed using 10 neural network trainings to distinguish BB1 from morphed R\&D.}
\end{table}

\section{Conclusion}
In this paper, we have presented a normalizing flow-based implementation of the ``chained quantile morphing" (CQM) technique for correcting Monte Carlo simulations to better match experimental data. CQM matches the performance of reweighting with neural networks and comes with the added benefit of producing a \textit{new set of corrected samples} rather than simply event weights. Moreover, quantile mapping is a fundamentally different approach to reweighting and can be particularly effective when correcting for detector mis-modeling. 

Our approach is unique in that the chained structure makes the implementation and interpretation of the results markedly simple and robust. The iterative, conditional density estimation -- as opposed to a simultaneous \textit{joint} estimation -- allows intervention at each stage of the morphing process, where the flow architectures can be modified or re-trained to maximize performance. This means we can ensure a high-quality fit to each variable without re-training the entire morphing process for every small modification to the flow. The structure of CQM also ensures some degree of \textit{locality} in the morphing transformations since each variable is transformed according to 1D conditional CDFs. This is the optimal transport map for 1D data. When chained together for an $N$-dimensional problem, it is intuitive that each variable is not moved ``too far" from its initial value. This stands in contrast to the joint density approaches where the training ensures that the \textit{overall} base density is transformed to match the target but does not guarantee small overall corrections per variable.

In the emerging landscape of morphing and reweighting strategies, most applications have been focused on background estimation for resonant anomaly searches. While this is a worthwhile application, we propose broadening the discussion to consider how CQM -- and MC correction strategies in general -- can impact the LHC physics program. As seen in this study and experimentally in Ref.~\cite{qmorphcms}, CQM is a promising tool for reducing systematic uncertainties by correcting mis-modeled simulations. This is useful for a wide range of physics analyses using $\mathcal{O}(10)$ high-level inputs, especially those that rely on neural networks or boosted decision trees to build a classifier. We have also demonstrated CQM's potential for problems with very high-dimensional inputs (e.g.\ jet taggers), where contrastive learning allows us to compress and correct the relevant information in a low-dimensional feature space.

As our machine learning tools become more sophisticated, they will continue to expose and amplify the flaws in our particle physics simulations. The LHC physics program already relies heavily on ML tools, and we will inevitably come to a point where the uncertainties associated with a simulation-trained model significantly limit the sensitivity of a search or the precision of a measurement. CQM is not a perfect solution to this problem, but it represents a promising step towards the ultimate goal of robust and unbiased ML. Moving ahead, we hope to apply a version of CQM to real LHC data and look forward to the ongoing development and refinement of morphing strategies. 

\begin{acknowledgments}
We thank Gregor Kasieczka for helpful and constructive feedback on the manuscript.

PH, PM, and SR are supported in part by the Institute for Artificial Intelligence and Fundamental Interactions (IAIFI) through NSF Grant No. PHY-2019786, and the NSF Institute for AI Accelerated Algorithms for Data-Driven Discovery (NSF Grant \#PHY-2117997). Additional support comes from the FAIR Data program of the U.S. DOE, Office of Science, Advanced Scientific Computing Research, under contract number DE-AC02-06CH11357, and a DOE early career award. 
\end{acknowledgments}

\appendix

\section{The limits of morphing \& reweighting\label{ap:reweight}}

Both chained quantile morphing and reweighting are subject to different intrinsic limitations, which must inform the choice of which technique to use in a real-world application. Conceptually the style of applications are different between the two algorithms. Morphing corresponds to a shift of the simulated distribution to match the data. This  corresponds to a detector mis-modeling where the detector response or resolution are off and the distribution requires a shifting and scaling to match the observed resolution. A reweighting is an adjustment of the simulation to emphasize/de-emphaise regions of the generated parameter space. This corresponds to correction to the simulation to match the intrinsic generated parameter to reality. This intuitively corresponds to generation mis-modeling such as a missing higher order correction or shower parameter. As a result, with each there are both advantages and disadvantages.   

We have constructed two toy examples of situations where one approach may be preferred over the other:

\subsection{The limitations of reweighting}

One of the major strengths of a morphing scheme is that it is capable of correcting a distribution into a range not well-covered by the original simulated sample. In our toy example, we consider a 5-dimensional dataset where the ``data'' has a strong correlation between two of the dimensions that is missed in the ``MC'' sample. In particular, we define the distributions
\begin{equation}
    v_i^{(MC)} = \mathcal{N}(0,1) \,\,\forall\,\, i
\end{equation}
\begin{align}
    v_i^{(data)} &= \mathcal{N}(0,1) \,\,\forall \,\,i\neq2 \\
    v_2^{(data)} &= \mathcal{N}(0,1) + 5v_3
\end{align}
Note that this creates a situation in which the ''MC'' sample is not completely overlapping with the ``data'' sample. 

\begin{figure}[h]
    \centering
    \includegraphics[width=0.4\textwidth]{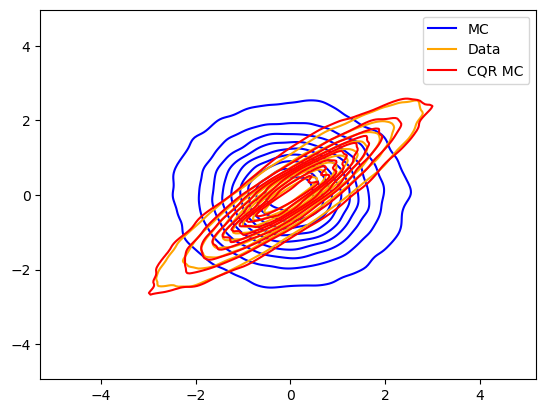}
    \includegraphics[width=0.4\textwidth]{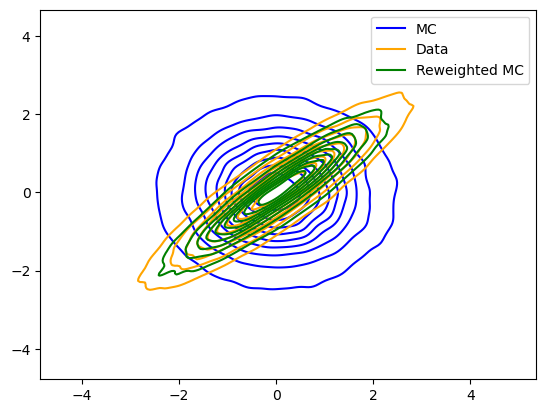}
    \caption{Kernel density estimate contours for ''MC'' and ''data'' distributions for our toy example, together with the CQR-corrected distribution (\textbf{top}) and the reweighting-corrected distribution (\textbf{bottom}).}
    \label{fig:A.1}
\end{figure}

We then corrected the MC distribution onto the data distribution with both CQR and a simple reweighting scheme. Kernel density estimate contours of the resulting distribution are shown in Figure \ref{fig:A.1}, where it is clear that the reweighting scheme fails to reproduce the data distribution where it fails to overlap the MC distribution. This would likely be improved with a larger dataset where the reweighting has more examples in the MC tails to reweight into the data distribution. In general, we find the CQR scheme to be more robust than reweighting, particularly in matching parts of the distribution that are not well-covered by the source distribution. 

\subsection{The limitations of morphing}

On the other hand, one of the major strengths of reweighting is that it preserves \textit{by construction} all of the complicated relationships between the variables in any given event. In situations where the variables being considered are related by some invariant or conserved quantity reweighting guarantees that that relationship is preserved in the transformed distribution. Morphing makes no such guarantee, and only preserves this relationship insofar as it is able to learn this relationship in the training. 

\begin{figure}[h]
    \centering
    \includegraphics[width=0.4\textwidth]{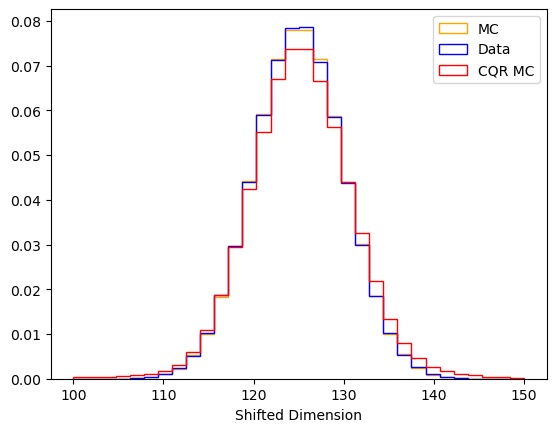}
    \includegraphics[width=0.4\textwidth]{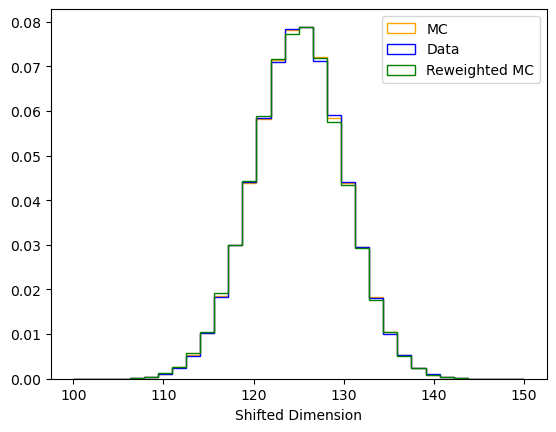}
    \caption{Reconstructed invariant mass distributions for ''MC'' and ''data'' distributions for our toy example, together with the CQR-corrected distribution (\textbf{top}) and the reweighting-corrected distribution (\textbf{bottom}).}
    \label{fig:A.2}
\end{figure}

In order to demonstrate this effect we consider a simplified toy model of the decay of a narrow resonance (suggestively placed at 125 GeV). In this dataset we simulate the decay of this resonance by first pulling an invariant mass from a normal distribution $\mathcal{N}(125, 5)$. One decay daughter is generated by randomly generating an momentum and angle in the 2D plane, and the second daughter is generated by randomly generating a $\delta \phi$ and then fixing the momentum to perfectly conserve the generated invariant mass. This creates a complex and perfectly fixed relationship between the momentum of the second particle and the momentum of the first particle, the $\delta \phi$ between them, and the generated invariant mass. In the ``MC'' sample the momentum distribution and $\delta \phi$ distribution are perturbed by $\approx 5\%$ with respect to the ``data'' sample, but the invariant masses are drawn from exactly the same distribution. We then blind the generated invariant mass and use both reweighting and CQR to correct the ``MC'' distributions onto the ``data'' ones. In order to evaluate the performance of the two schemes at preserving the invariant mass we reconstructed it from the two decay daughters and show the resulting distributions in Figure \ref{fig:A.2}. The reweighting scheme perfectly preserves the invariant mass, while the CQR isn't able to perfectly learn this relationship and smears out the invariant mass distribution. This could likely be improved with more data or more fine-tuning of the trainings in order to help the CQR model learn the complex relationship between the variables. An alternative approach would be to reconstruct the invariant mass distribution in both the data and MC samples first, and then perform a morphing of this distribution from MC to data.

% The \nocite command causes all entries in a bibliography to be printed out
% whether or not they are actually referenced in the text. This is appropriate
% for the sample file to show the different styles of references, but authors
% most likely will not want to use it.
%\nocite{*}
\bibliography{refs}% Produces the bibliography via BibTeX.

\end{document}